\documentclass[12pt,english]{article}
\usepackage[T1]{fontenc}
\usepackage[latin9]{inputenc}
\usepackage{geometry}
\usepackage{babel}
\usepackage{float}
\usepackage{amsmath}
\usepackage{graphicx}
\usepackage{setspace}
\usepackage[authoryear]{natbib}
\usepackage[unicode=true,pdfusetitle,
 bookmarks=true,bookmarksnumbered=false,bookmarksopen=false,
 breaklinks=false,pdfborder={0 0 0},pdfborderstyle={},backref=false,colorlinks=false]
 {hyperref}

\makeatletter
\usepackage{enumitem}
\setlist{nolistsep}

\usepackage{setspace}
\usepackage{lineno}
\usepackage{natbib}
\usepackage{pdflscape}
\usepackage{hyperref}
\usepackage{multirow}

\usepackage{titlesec}
\titlespacing*{\subsubsection}{0pt}{1.0ex}{0pt}
\titlespacing*{\subsection}{0pt}{1.0ex}{0pt}

\g@addto@macro\normalsize{%
  \setlength\abovedisplayskip{2pt}
  \setlength\belowdisplayskip{2pt}
  \setlength\abovedisplayshortskip{1pt}
  \setlength\belowdisplayshortskip{1pt}
}

\makeatother

\begin{document}
\begin{flushleft}
\setcounter{footnote}{0}
\begin{spacing}{2.0}
\begin{flushleft}Running Head: Spatial occupancy models
\par\end{flushleft}

\noindent \textbf{Using machine learning to identify nontraditional
spatial dependence in occupancy data}

\medskip{}

\begin{doublespace}
\textbf{Narmadha M. Mohankumar}\footnote{Corresponding author; E-mail: meenu@ksu.edu.}

{\small{}Department of Statistics, Kansas State University}{\small\par}

\medskip{}

\textbf{Trevor J. Hefley}

{\small{}Department of Statistics, Kansas State University}\medskip{}

\end{doublespace}

\textbf{Open Research statement: }The Thomson's gazelle data and sugar
glider data used in our data example both are available from the Dryad
Digital Repository (\citet{Thompsonsgazelle,sugargliderdata}): https://doi.org/10.5061/dryad.34kb373
and https://doi.org/10.5061/dryad.xgxd254bt, respectively. A tutorial
showing how to implement our statistical model is provided in Appendix
S1. Annotated computer code that can be used to reproduce all results
and figures associated with the simulation experiment and data examples
are provided in Appendix S2, S3, and S4. 

\begin{doublespace}
\bigskip{}

\newpage{}
\end{doublespace}

\setlength{\parindent}{0.7cm}
\begin{doublespace}

\subsection*{Abstract\vspace{0.2cm}
}
\end{doublespace}

\begin{doublespace}
\noindent  Spatial models for occupancy data are used to estimate
and map the true presence of a species, which may depend on biotic
and abiotic factors as well as spatial autocorrelation. Traditionally
researchers have accounted for spatial autocorrelation in occupancy
data by using a correlated normally distributed site-level random
effect, which might be incapable of identifying nontraditional spatial
dependence such as discontinuities and abrupt transitions. Machine
learning approaches have the potential to identify and model nontraditional
spatial dependence, but these approaches do not account for observer
errors such as false absences. By combining the flexibility of Bayesian
hierarchal modeling and machine learning approaches, we present a
general framework to model occupancy data that accounts for both traditional
and nontraditional spatial dependence as well as false absences. We
demonstrate our framework using six synthetic occupancy data sets
and two real data sets. Our results demonstrate how to identify and
model both traditional and nontraditional spatial dependence in occupancy
data which enables a broader class of spatial occupancy models that
can be used to improve predictive accuracy and model adequacy.

\noindent \textit{Key-words: hierarchical Bayesian model; machine
learning; occupancy model; presence--absence data; site occupancy;
spatial dependence; zero-inflated binomial model}\vspace{-0.1cm}

\end{doublespace}

\noindent 

\subsection*{Introduction}

\begin{doublespace}
Many ecological studies collect occupancy data to understand the dynamics
of species occurrence over space and time (e.g., \citealt{hepler2018identifying,joseph2020neural}).
 Occupancy data are collected by making replicated visits to sites
and recording the presence or absence of at least one individual.
During a site visit, individuals may go undetected even when present,
resulting in the detection of no individuals (i.e., a false absence).
Failure to account for false absences can have a significant impact
on parameter estimates and predictions (\citealt{hoeting2000improved,mackenzie2002estimating,tyre2003improving}).

To facilitate the analysis of occupancy data that contain false absences,
\citet{hoeting2000improved}, \citet{mackenzie2002estimating}, and
\citet{tyre2003improving} introduced a zero-inflated Bernoulli model
that specifies a distribution of the observed data given the true
 presence at a site. Using familiar notation for Bayesian hierarchical
models, the conditional distribution of the data is
\begin{equation}
y_{ij}|z_{i},p_{ij}\sim\left\{ \begin{matrix}\text{Bernoulli}(p_{ij}) & ,z_{i}=1\\
0 & ,z_{i}=0
\end{matrix}\right.,
\end{equation}
where $y_{ij}=1$ denotes the presence and detection of one or more
individuals at the $i^{\text{th}}$ site ($i=1,2,...,n$) during the
$j^{\text{th}}$ sampling period $(j=1,2,...,J_{i})$ and $y_{ij}=0$
denotes that no individuals were detected. Detection of at least one
individual depends on the probability $p_{ij}$. The $z_{i}$ is the
true presence ($z_{i}=1$) or absence ($z_{i}=0$) at the $i^{\text{th}}$
site, which is assumed to be constant during all $J_{i}$ sampling
periods and modeled as
\begin{equation}
z_{i}|\psi_{i}\sim\text{Bernoulli}(\psi_{i})\;.
\end{equation}
In Eq. 2, the probability of true presence, $\psi_{i}$, is modeled
using an intercept term and $q$ site-level covariates with the equation
\begin{equation}
g^{-1}(\psi_{i})=\mathbf{x}_{i}^{'}\boldsymbol{\beta}\;,
\end{equation}
where $g^{-1}(\cdot)$ is an appropriate link function (e.g., logit
or probit), $\mathbf{x}_{i}\equiv(1,x_{1},x_{2},...,x_{q})^{'}$,
and $\boldsymbol{\beta}\equiv(\beta_{0},\beta_{1},\beta_{2},...,\beta_{q})^{'}$.
Within the vector $\boldsymbol{\beta}$, $\beta_{0}$ is the intercept
parameter and $\beta_{1},\beta_{2},...,\beta_{q}$ are regression
coefficients. 

Since the introduction of the occupancy model in Eqs. 1--3, many
extensions were developed to address model inadequacies. For example,
to account for spatial dependence \citet{johnson2013spatial} added
a correlated normally distributed site-level effect, $\eta_{i}$  (i.e.,
$(\eta_{1},\eta_{2},...,\eta_{n})^{'}\sim\text{N}(\mathbf{0},\boldsymbol{\Sigma})$;
see ch. 26 in \citealt{hooten2019bringing}) to Eq. 3 that resulting
in
\begin{equation}
g^{-1}(\psi_{i})=\mathbf{x}_{i}^{'}\boldsymbol{\beta}+\eta_{i}\;.
\end{equation}

\end{doublespace}

\begin{doublespace}
\noindent The approach by \citet{johnson2013spatial} has been effective
in accounting for occupancy model inadequacies caused by traditional
spatial dependence (e.g., \citealt{wright2019identifying}), which
is assumed to have been generated from a correlated normally distributed
random effect that imparts varying levels of smoothness on the spatial
process. Discontinuities, abrupt transitions, and other ``non-normal''
spatial processes are common in ecological data, and the traditional
spatial random effect may fail to capture such dynamics (e.g., \citealt{hefley2017dynamic}).
Unfortunately, ecologists lack alternative occupancy model specifications
that would allow them to check for and, if needed, model nontraditional
spatial dependence.
\end{doublespace}

\begin{doublespace}
We demonstrate a framework for occupancy data to identify and model
both traditional and nontraditional spatial dependence. Our framework
takes a machine learning approach to model the site-level effect in
Eq. 4 and can identify both traditional and nontraditional spatial
dependence. We illustrate this framework using six synthetic data
sets containing traditional and nontraditional spatial dependence
and  then apply our approach to understand the spatial dynamics of
Thomson\textquoteright s gazelle (\textit{Eudorcas thomsonii}) in
Tanzania and sugar gliders (\textit{Petaurus breviceps}) in Tasmania.

\vspace{-1.4cm}

\end{doublespace}

\subsection*{Materials and Methods\vspace{-0.3cm}
}

\subsubsection*{\textmd{\textit{Occupancy data requirements}}}

\begin{doublespace}
Our proposed modeling framework builds upon the occupancy model of
\citet{mackenzie2002estimating} and \citet{tyre2003improving} and
therefore is intended for use with occupancy data that was collected
with repeated site visits during which the true presence or absence
of individuals at a site does not change. In addition, we require
that false negative detections are the only observational error. However,
our framework is adaptable to accommodate other types of occupancy
data (see ``Model extensions'' in Appendix S1 for additional detail).
For example, our framework can be adapted to account for false presence,
which occurs when individuals are not present at a site but are recorded
as occurring at a site.

\vspace{-1.1cm}

\end{doublespace}

\subsubsection*{\textmd{\textit{Spatial occupancy model framework}}}

\begin{doublespace}
Our proposed framework involves lifting the normal distributional
assumption in the spatial component that accounts for the spatial
dependence. To accomplish this, we replace the site-level effect in
Eq. 4 with 
\begin{equation}
g^{-1}(\psi_{i})=\mathbf{x}_{i}^{'}\boldsymbol{\beta}+f(\mathbf{s}_{i})\;.
\end{equation}
Conceptually, this is an important change; the $f(\mathbf{s}_{i})$
is an unknown spatially varying process that is a function, $f(\cdot)$,
that depends on the coordinate vector, $\mathbf{s}_{i}$, of the $i^{\text{th}}$
site. The function $f(\cdot)$ is always unknown and is approximated. 

This change in perspective is common in the field of machine learning,
where the goal is to ``learn'' or approximate an underlying function
using data (see ch. 5 in \citealt{hastie2009elements}). This simple
change in Eq. 5  expands the types of model specifications for the
spatially varying process, $f(\mathbf{s}_{i})$. For example, regression
trees are used to learn about underlying functions that have discontinuities
and abrupt transitions, and using regression trees to approximate
$f(\mathbf{s}_{i})$ could model nontraditional spatial dependence. 

Many approaches from machine learning, such as support vector regression,
neural networks, boosted regression trees, and Gaussian processes,
could approximate $f(\cdot)$. These approaches have been widely used
by ecologists to make predictions and inferences about species distributions
from abundance and presence-absence data (e.g., \citealt{de2000classification,cutler2007random,elith2008working,golding2016fast}).
However, machine learning approaches are not widely used to model
occupancy data because of the issues associated with false absences.
Furthermore, approximating the spatial dependence within the occupancy
model using machine learning approaches requires custom programming
and a level of technical knowledge that hinders widespread use. The
existing approaches that blend machine learning approaches with occupancy
models are approach specific (e.g., \citealp{hutchinson2011incorporating,joseph2020neural}),
and therefore switching among the different types of approaches to
approximate $f(\cdot)$ is a challenge. For example, switching from
a neural network to a regression tree to approximate $f(\cdot)$ in
Eq. 5 would require extensive retooling of computer code, thus hindering
model checking, comparisons, and selection. 

Fortunately, \citet{shaby2012embedding} developed a model-fitting
algorithm based on Markov chain Monte Carlo (MCMC) that enables off-the-shelf
software for machine learning approaches, such as those available
in R (e.g., \texttt{rpart(...)}, \texttt{svm(...)}, \texttt{gam(...)}),
to be embedded within hierarchical Bayesian models. Once the initial
computer code is written for the occupancy model, switching among
machine learning approaches to approximate $f(\cdot)$ requires modifying
only a few lines of code. Details associated with model fitting are
provided in Appendix S1 of the Supplementary Material.

\vspace{-1cm}

\end{doublespace}
\begin{doublespace}

\subsubsection*{\textmd{\textit{Identifying spatial dependence}}\vspace{0.2cm}
}
\end{doublespace}

\begin{doublespace}
To identify the type of spatial dependence and check model adequacy,
we use a model selection and model checking approach. First, we use
a wide variety of approaches to model spatial dependence and then
use a measure of predictive accuracy to determine which approach most
accurately models the spatial process. We supplement this predictive
approach with a measure of model adequacy (e.g., \citealt{wright2019identifying}).

Following \citet{hobbs2015bayesian}, we measure the predictive accuracy
using $-2\times\text{LPPD}$, where LPPD is the log posterior predictive
density. The $-2\times\text{LPPD}$ is similar to the information
criterion used for model selection but uses out-of-sample data rather
than in-sample data (\citealt{hooten2015guide}). As such, $-2\times\text{LPPD}$
and the difference in $-2\times\text{LPPD}$ among models can be interpreted
similarly to the information criterion that attempts to approximate
$-2\times\text{LPPD}$ using in-sample data (e.g., Watanabe-Akaike
information criteria). For example, if model A produced a $-2\times\text{LPPD}$
score less than the $-2\times\text{LPPD}$ score produced by model
B, then model A has higher predictive accuracy. As a standard of comparison,
we fit an occupancy model that does not account for spatial dependence
(i.e., Eq. 3; hereafter nonspatial occupancy model).

In addition, we use Moran's I correlogram to check model adequacy.
Moran's I has been used to detect traditional spatial dependence in
the residuals of fitted occupancy models (\citealt{wright2019identifying}).
However, if traditional approaches fail to capture spatial dependence,
then Moran's I may identify such inadequacies. 

\vspace{-1cm}

\end{doublespace}
\begin{doublespace}

\subsubsection*{\textmd{\textit{Synthetic data examples}}\vspace{0.1cm}
}
\end{doublespace}

\begin{doublespace}
\noindent For our synthetic data examples, we show the probability
of occupancy in Fig. 1, which includes the three scenarios of nontraditional
spatial dependence and the three scenarios of traditional spatial
dependence listed below.\vspace{-0.8cm}

\end{doublespace}
\begin{enumerate}
\begin{doublespace}
\item Spatial dependence that has discontinuities and abrupt transitions
generated by a step-wise function (nontraditional; Fig. 1a).
\item Spatial dependence forming a circle with the probability of occupancy
being low in the center and smoothly increases towards the edges (nontraditional;
Fig. 1b).
\item Spatial dependence defined by a cosine function (nontraditional; Fig.
1c).
\item Normally distributed random effect with a correlation matrix specified
by a conditional autoregressive process (traditional; Fig. 1d).
\item Normally distributed random effect with a correlation matrix specified
by an exponential covariance function (traditional; Fig. 1e).
\item Normally distributed random effect with a correlation matrix specified
by a squared exponential covariance function (traditional; Fig. 1f).
\end{doublespace}
\end{enumerate}
\vspace{-0.8cm}

\begin{doublespace}
For each scenario, we generate synthetic data using Eqs. 1, 2, and
5 on a unit square study area (i.e., $\mathcal{S=}[0,1]\ensuremath{\times}[0,1]$).
We divided the study area, $\mathcal{S}$, into 900 grid cells (sites).
We set the true values for the parameters to $p_{ij}=0.5$ and $\beta_{0}=0$
. We exclude covariates and regression coefficients in our synthetic
data so that the spatial process is unobstructed when $\psi_{i}$
is mapped onto $\mathcal{S}$, which aids when visual and numerical
comparisons are made among the machine learning approaches. From the
900 grid cells, we consider a random sample of $n=200$ sites as the
study area with $J_{i}=3$ visits.

We apply our spatial occupancy modeling framework to the six synthetic
data sets and compare the performance of four embedded machine learning
approaches, which include regression trees, support vector regression,
a low-rank Gaussian process, and a Gaussian Markov random field. The
low-rank Gaussian process and Gaussian Markov random field are approaches
that model traditional spatial dependence for data sets with a large
number of sites and have been used in models for occupancy data (\citealt{johnson2013spatial,heaton2019case}).
The regression tree and support vector regression are nontraditional
approaches and may be capable of identifying nontraditional types
of spatial dependence.

We assess the performance of each approach to identify spatial dependence
using $-2\times\text{LPPD}$ calculated at 200 sites that were not
used for model fitting (hereafter out-of-sample sites) and by using
Moran's I correlogram. In addition, we visually compare the true probability
of occupancy ($\psi_{i}$) to the posterior mean of the probability
of occupancy ($\text{E}($$\psi_{i}|\mathbf{y})$; Fig. 2). Details
associated with the synthetic data are provided in Appendix S2 of
the Supplementary Material. 

\vspace{-1.1cm}

\end{doublespace}

\subsubsection*{\textmd{\textit{Thomson's gazelle data}}}

\begin{doublespace}
We illustrate our spatial occupancy modeling framework using a data
set from \citet{hepler2018identifying}, who reported the presence
and absence of Thomson's gazelle at 195 sites within Serengeti National
Park, Tanzania (Fig. 3a). The sites were sampled using a network of
motion-sensitive and thermally activated cameras. Images were classified
by participants on the citizen science website Snapshot Serengeti.
A site visit consisted of an 8-day period during the year 2012 (e.g.,
January 1--8, 2012). Each site was visited between 1 and 46 times
(the mean number of visits was 29). Following \citet{hepler2018identifying},
$y_{ij}=1$ (from Eq. 1) was recorded if an image of at least one
Thomson\textquoteright s gazelle was captured at the $i^{\text{th}}$
site within the $j^{\text{th}}$ 8-d window. A value of $y_{ij}=0$
was recorded if the site was sampled, but no individuals were observed.
Of the 195 sites, 141 had at least one detection. We use 100 randomly
selected sites for model fitting and reserve the remaining 95 sites
to calculate $-2\times\text{LPPD}$.

Similar to our synthetic data example, we apply our spatial occupancy
modeling framework by embedding four machine learning approaches,
which include regression trees, support vector regression, a low-rank
Gaussian process, and a Gaussian Markov random field. We exclude site-level
covariates in our data example to illustrate our approaches ability
to model multiple processes that generate spatial dependence (e.g.,
missing site-level covariates and spatial autocorrelation) and to
illustrate the ability of our method to serve as a ``spatial interpolator''
for occupancy data (i.e., similar to indicator or binomial kriging,
but accounting for false absences). However, as with traditional occupancy
models, we can easily include site-level covariates into our spatial
occupancy models.  Details associated with  the data example are provided
in Appendix S3 of the Supplementary Material.\vspace{-1.1cm}

\end{doublespace}

\subsubsection*{\textmd{\textit{Sugar glider data}}}

\begin{doublespace}
We illustrate our modeling framework using a second data set from
\citet{allen2018occupancy}, who reported the presence and absence
of sugar gliders. The data were collected during four or five site
visits made to 100 sites in the Southern Forest region of Tasmania
(Fig. 4a). Of the 100 sites, 79 had at least one sugar glider detected.
Because this data set has a relatively small number of sites, we used
75 randomly selected sites for model fitting and reserve the remaining
25 sites to calculate $-2\times\text{LPPD}$. We use the same modeling
approaches for this example as we did in the Thomson's gazelle example.
Details associated with the data example are provided in Appendix
S4 of the Supplementary Material.

\vspace{-1.4cm}

\end{doublespace}

\subsection*{Results\vspace{-0.3cm}
}

\subsubsection*{\textmd{\textit{Synthetic data examples}}}

\begin{doublespace}
In scenario 1,  the occupancy model with an embedded regression tree
performed best because the other embedded machine learning approaches
didn't capture the abrupt transition created by the step-wise spatial
process (Fig. 2). The $-2\times\text{LPPD}$ was $348.5$, $377.2$,
$377.5$, and $384.0$ for the embedded regression tree, support vector
regression, low-rank Gaussian process, and Gaussian Markov random
field, respectively. For comparison, the $-2\times\text{LPPD}$ obtained
from the nonspatial occupancy model was 433.1. Similarly, for scenario
1, the comparison of the Moran's I between the occupancy models suggested
that spatial dependence must be accounted for using a regression tree;
all other approaches resulted in lingering spatial dependence (see
Fig. S6 in Appendix S5).

Detailed results for scenarios 2--6 are presented in Appendix S5
of the Supplementary Material. For example, in scenario 2, the spatial
dependence forms a circle with the probability of occupancy being
low in the center and smoothly increases towards the edge of the circle
(Fig. 1b). For scenario 2, we expected and found that the embedded
support vector regression performed best (see Appendix S5). This was
expected because this machine learning approach is best suited to
learn about smoothly varying deterministic functions. In total, the
results from the scenarios clearly demonstrated that if the spatial
process is a discontinuous step function, then the approaches used
to model traditional spatial dependence are not adequate, and the
approaches such as regression trees should be used. If the spatial
dependence is traditional, the differences among the approaches are
less distinct; nevertheless, in general, support vector regression
performs superior for smoothly varying processes (see Appendix S5).

\vspace{-1.2cm}

\end{doublespace}

\subsubsection*{\textmd{\textit{Spatial occupancy dynamics of Thomson's gazelle}}}

\begin{doublespace}
Across the four embedded machine learning approaches, the probability
of occupancy at a site ranged from 0.45 to 0.95 (Fig. 3b--3e). Generally,
the probability of occupancy was high across the entire study area.
However, there was a distinct band running from the southwest to the
northeast of the study area where the probability of occupancy was
much lower (Fig. 3b--3e). 

The measure of predictive accuracy, $-2\times\text{LPPD}$, was $669.4$,
$668.8$, $671.0$, and $668.7$ for embedded regression trees, support
vector regression, a low-rank Gaussian process, and a Gaussian Markov
random field, respectively. For comparison, the $-2\times\text{LPPD}$
obtained from the non-spatial occupancy model was 676.7. Comparison
of the Moran's I between the non-spatial and spatial occupancy models
suggested that accounting for spatial dependence improves model adequacy;
although, the utility of Moran's I is questionable because the differences
among approaches are trivial, which may be due to the small number
of sites (see Fig. S1 in Appendix S3; \citealt{carrijo2017modified}).
In total, the $-2\times\text{LPPD}$ and Moran's I suggest that spatial
dependence should be accounted for in the model. However, Moran's
I and $-2\times\text{LPPD}$ suggested that the differences among
machine learning approaches are less distinct; therefore, it is unclear
if the spatial dependence is traditional or nontraditional. 

\vspace{-1.2cm}

\end{doublespace}

\subsubsection*{\textmd{\textit{Spatial occupancy dynamics of sugar gliders}}}

\begin{doublespace}
For the sugar glider data example, the probability of occupancy at
a site ranged from 0.48 to 0.97 (Fig. 4b--4e) across the four embedded
machine learning approaches. The probability of occupancy was generally
high across the entire study area; however, there was an area in the
eastern and southeastern portion of the study area where the probability
of occupancy was relatively low (i.e., $\psi_{i}<0.60)$, and there
were clear visual differences in the probability of occupancy among
the four machine learning approaches (Fig. 4b--4e). The measure of
predictive accuracy, $-2\times\text{LPPD}$, was $78.2$, $80.4$,
$79.6$, and $78.9$ for embedded regression trees, support vector
regression, a low-rank Gaussian process, and a Gaussian Markov random
field, respectively. For comparison, the $-2\times\text{LPPD}$ obtained
from the non-spatial occupancy model was 80.3. Similar to Thomson's
gazelle example, the comparison of the Moran's I between the occupancy
models suggested that accounting for spatial dependence improves model
adequacy (see Fig. S1 in Appendix S4). In total, the $-2\times\text{LPPD}$
and Moran's I suggest that  the spatial process (i.e., $f(\cdot)$
in Eq. 5) is best modeled using a regression tree. Using Moran's I
and $-2\times\text{LPPD}$ as evidence, the results suggested that
the spatial dependence is nontraditional. 

\vspace{-1.3cm}

\end{doublespace}

\subsection*{Discussion}

\begin{doublespace}
The use of occupancy models has increased rapidly since the early
2000s. Occupancy data are inherently spatial, but unfortunately, only
a limited number of approaches existed to model the spatial process
(i.e., \citealt{hoeting2000improved,johnson2013spatial}). This lack
of spatial modeling options for occupancy data is in contrast to species
distribution models (SDM) that predict the spatial distribution of
a species using statistical and machine learning approaches applied
to presence-only, count, and presence-absence data. There is a bewildering
number of approaches within the SDM literature that are used to model
the spatial process. Unfortunately, many of the SDM approaches do
not account for contamination in the response variable (e.g., false
absences). Understandably ecologists may feel forced to choose between
SDM approaches that do not account for contamination in the response
variable (e.g.,  regression trees) and approaches that do, but with
a lack of spatial modeling (e.g., occupancy models).

The crux for ecologists planning to use our framework is to determine
which machine learning approaches are likely to capture the spatial
process, which will require a level of familiarity with the properties
of a wide range of machine learning approaches. We recommend \citet{james2013introduction}
for a gentle introduction and \citet{hastie2009elements} and \citet{murphy2012machine}
for more advanced and broad presentations. Within the ecological literature,
there are also several excellent guides to machine learning approaches
(e.g., \citealt{de2000classification,cutler2007random,elith2008working}).

Recently, the hierarchical modeling framework commonly used in ecology
has been expanded to include some types of machine learning approaches
such as neural networks (\citealt{wikle2019comparison,joseph2020neural}).
Our study builds upon this previous work and expands the types of
spatial models ecologists can use for data that fit within the occupancy
model framework. Although our work is focused on spatial dependence
among the true presence at a site, the approach is easily generalizable.
For example, Eq. 5 implies a linear effect of the site-level covariates
(i.e., $\mathbf{x}_{i}^{'}\boldsymbol{\beta}$). \citet{shaby2012embedding}
show how machine learning approaches can be used to capture nonlinear
and unknown relationships between covariates and the probability of
occupancy, thus alleviating the linear assumption in Eq. 5. Furthermore,
many studies that use occupancy models perform covariate selection
using model selection techniques (e.g., \citet{hooten2015guide}).
While model selection techniques work for a small number of covariates,
machine learning approaches may be superior when there are a large
number of covariates. Another important generalization is that the
machine learning approaches can be embedded to model the probability
of detection as a function of predictor variables such as Julian date
and observer effort (i.e., similar to the use of cubic splines used
by \citet{johnston2018estimates}). To facilitate these extensions,
we explain in Appendix S1 how to generalize our framework for other
popular ecological models, which is a direct application of the work
by \citet{shaby2012embedding}. 

\vspace{-1.3cm}

\end{doublespace}

\subsection*{Acknowledgements}

We thank Drs. Staci Hepler, Michael Anderson, and Robert Erhardt for
assistance with the Thomson's gazelle data. We thank Dr. Daniel Fink,
an anonymous reviewer, and Dr. Julia Jones for comments that improved
our manuscript. Funding for this project was provided by the National
Science Foundation via grant DEB 1754491.

\renewcommand{\refname}{\vspace{-0.8cm}\selectfont\large Literature Cited\vspace{-0.4cm}}\setlength{\bibsep}{0pt}\setstretch{1.76}

\bibliographystyle{apa}
\bibliography{refs}
\vspace{-0.3cm}

\noindent 

\subsection*{\protect\pagebreak Figure legends\vspace{-0.1cm}
}

\noindent 

\noindent 

\noindent 

\noindent \textbf{Figure 1.} Synthetic data examples showing the probability
of occupancy ($\psi_{i}$ in Eq. 5) at 900 potential sites (pixels)
for six scenarios of traditional and nontraditional spatial dependence.
The nontraditional scenarios include spatial dependence having discontinuities
and abrupt transitions (panel a), forming a circle (panel b), and
defined by a cosine function (panel c). The traditional scenarios
include spatial dependence generated from a normally distributed random
effect with a correlation matrix specified using a conditional autoregressive
process (panel d), an exponential covariance function (panel e), and
a squared exponential covariance function (panel f). \vspace{-0.1cm}

\noindent 

\noindent \textbf{Figure 2.} The probability of occupancy from scenario
1 of the synthetic data example (panel a) and the posterior mean of
the probability of occupancy ($\text{E}(\psi_{i}|\mathbf{y})$)) obtained
by fitting spatial occupancy models that included an embedded regression
tree (panel b), a support vector regression (panel c), a low-rank
Gaussian process (panel d), and a Gaussian Markov random field (panel
e). The gray squares in the panel are the locations of the 200 sampled
sites used for model fitting.\vspace{-0.1cm}

\noindent \textbf{Figure 3. }Thomson's gazelle data from \citet{hepler2018identifying}
collected at 195 sites within Serengeti National Park, Tanzania (panel
a) and the posterior mean of the probability of occupancy ($\text{E}(\psi_{i}|\mathbf{y})$;
panels b--e). Panels b--e show $\text{E}(\psi_{i}|\mathbf{y})$
obtained by fitting spatial occupancy models that included an embedded
regression tree (panel b), a support vector regression (panel c),
a low-rank Gaussian process (panel d), and a Gaussian Markov random
field (panel e).\vspace{-0.1cm}

\noindent \textbf{Figure 4. }Sugar glider data from \citet{allen2018occupancy}
collected at 100 sites in the Southern Forest region of Tasmania and
the posterior mean of the probability of occupancy ($\text{E}(\psi_{i}|\mathbf{y})$;
panels b--e). Panels b--e show $\text{E}(\psi_{i}|\mathbf{y})$
obtained by fitting spatial occupancy models that included an embedded
regression tree (panel b), a support vector regression (panel c),
a low-rank Gaussian process (panel d), and a Gaussian Markov random
field (panel e).

\pagebreak\begin{landscape}
\begin{figure}[H]
\begin{centering}
\includegraphics[scale=1.3]{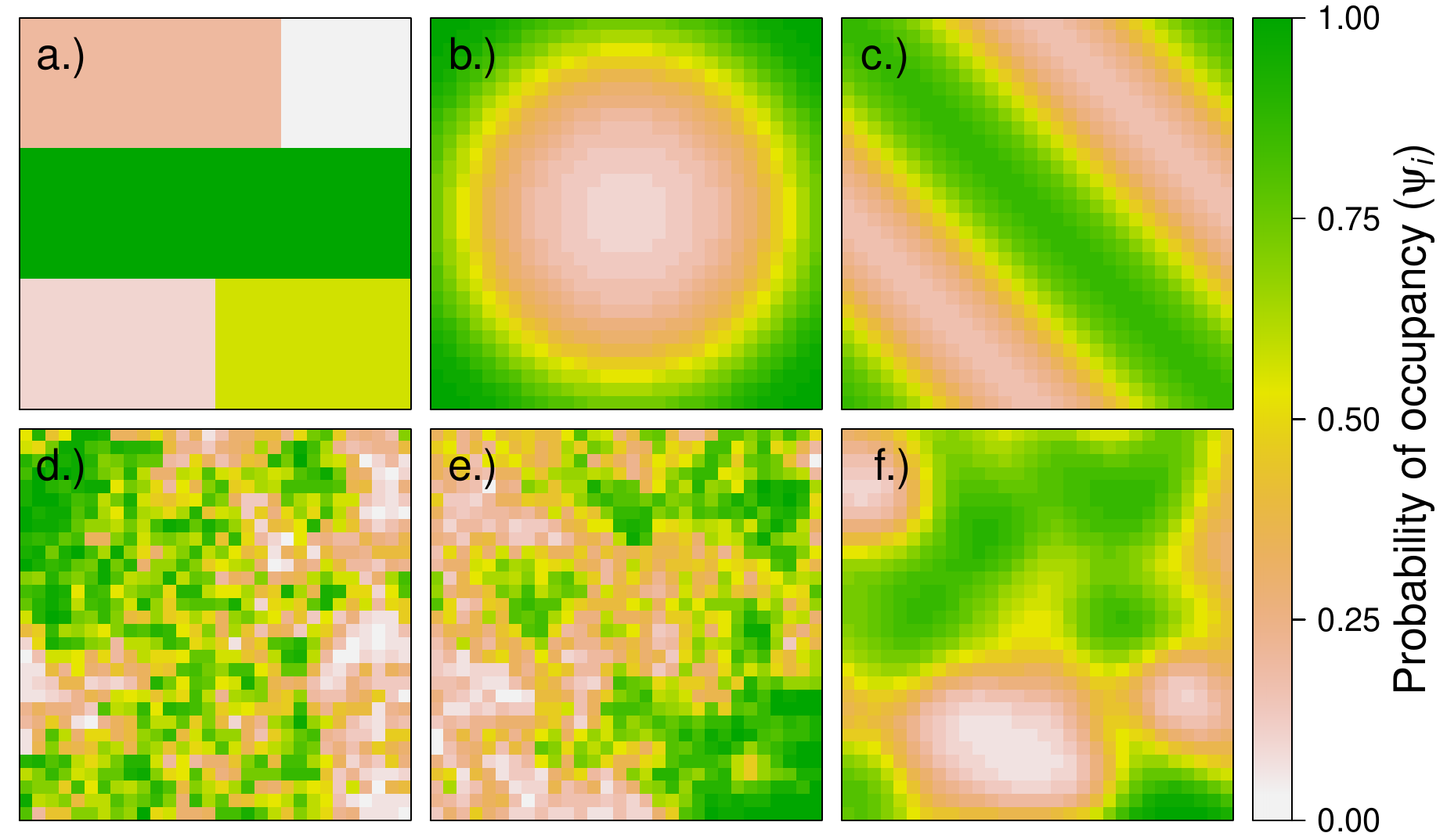}
\par\end{centering}
\caption{}
\end{figure}
\pagebreak{}
\begin{figure}[H]
\begin{centering}
\includegraphics[scale=0.95]{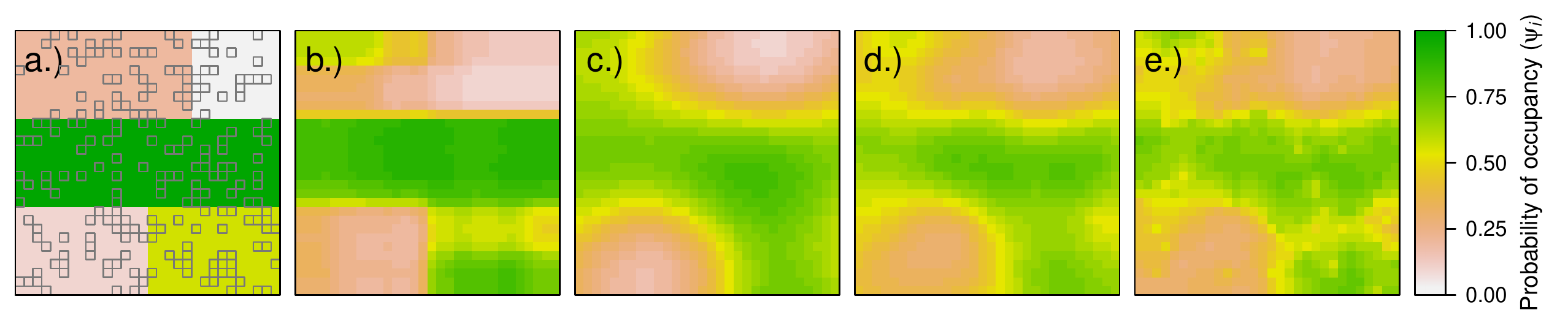}
\par\end{centering}
\caption{}
\end{figure}
\end{landscape}\pagebreak{}
\begin{figure}[H]
\begin{centering}
\includegraphics[scale=0.78]{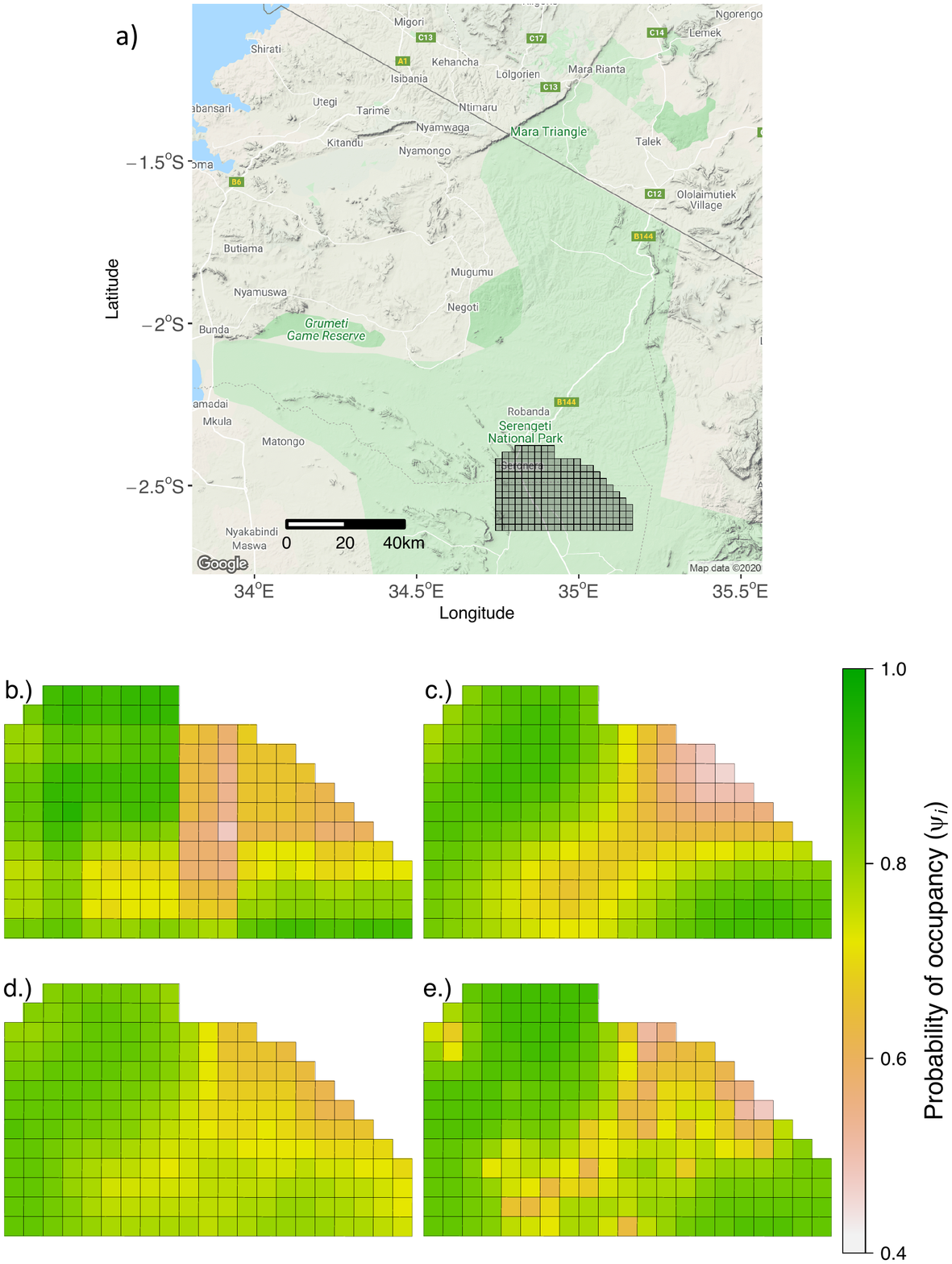}
\par\end{centering}
\centering{}\caption{}
\end{figure}
\pagebreak{}
\begin{figure}[H]
\begin{centering}
\includegraphics[scale=0.82]{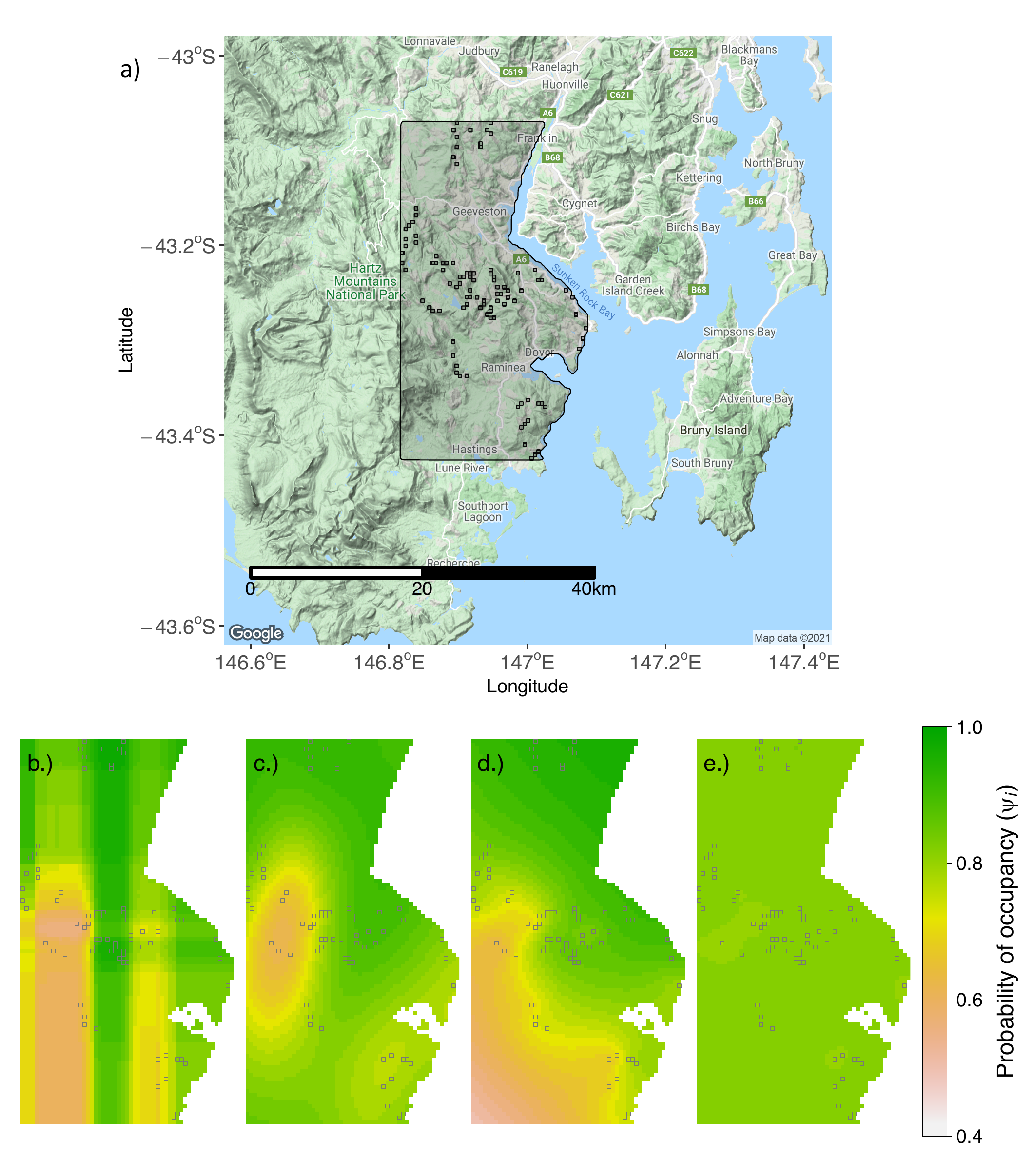}
\par\end{centering}
\centering{}\caption{}
\end{figure}

\end{spacing}
\end{flushleft}
\end{document}